# Pump-power-controlled L-band wavelength-tunable mode-locked fiber laser utilizing all polarization maintaining nonlinear polarization rotation

GUANYU YE,[1,2] BOWEN LIU,[1,2] MAOLIN DAI,[1,2] YIFAN MA,[1,2] TAKUMA SHIRAHATA,[1,2] SHINJI YAMASHITA,[1,2] SZE YUN SET[1, *]

[1] Research Center for Advanced Science and Technology, The University of Tokyo, 4-6-1 Komaba, Meguro-ku, Tokyo 153-8904, Japan
[2] Department of Electrical Engineering and Information Systems, The University of Tokyo, Bunkyo-ku, Tokyo 113-8656, Japan
*Corresponding author: set@cntp.t.u-tokyo.ac.jp

**For the first time, we present the pump power-controlled wavelength-tunable mode-locked fiber laser in the L-band (1565 nm to 1625 nm), achieved by all-polarization maintaining (all-PM) nonlinear polarization rotation (NPR). The wavelength of the laser can be tuned over 20 nm, from 1568.2 nm to 1588.9 nm simply by controlling the pump power from 45 mW to 115 mW. In contrast to conventional wavelength tuning mechanisms such as optical bandpass filters, our tuning method is non-mechanical and electrically controllable, featuring simplicity and cost-effectiveness in a superior all-fiber design.**

Wavelength-tunable mode-locked fiber lasers (MLFL) have attracted increasing attention, notably for their applications in various fields such as optical spectroscopy [1], bio-imaging [2], and optical sensing [3]. L-band (1565 nm to 1625 nm) tunable mode-locked lasers, capable of providing 800 nm optical pulses through second harmonic generation, offer a compact alternative to the bulky tunable Ti: Sapphire lasers used in microscopy applications [4, 5]. Various methods have been explored to achieve wavelength tunable mode-locking in the L-band. A common approach involves using external tunable optical bandpass filters such as fiber Bragg grating (FBG), chirped FBG, long-period fiber grating (LPFG), and Sagnac filter [6-9]. Other wavelength tuning methods have also been studied, with intracavity loss control using tunable-ratio optical couplers (TROC) or variable optical attenuators (VOA) [10, 11] to adjust the gain peak of the erbium-doped fiber (EDF) via the gain tilt effect, resulting in a tunable laser output wavelength [12]. Wavelength tuning using a polarization controller (PC) and a polarizer have also been studied [13, 14], where PCs are used to alter fiber birefringence and the intracavity polarization state to realize optical bandpass filtering effect for wavelength adjustment [15]. Furthermore, soliton self-frequency shift (SSFS) has also been utilized for wavelength tuning via the adjustment of the injected pulse energy [16].

However, these previously demonstrated wavelength tuning mode-locked lasers are implemented in non-polarization-maintaining (PM) configurations rendering them susceptible to environmental perturbations. The use of a PC in a fiber laser cavity further compromises laser reliability and repeatability, limiting practical applications.

In order to over these short-comings, recent demonstrations of robust and reliable wavelength tunable MLFLs uses all-PM-fiber setups featuring the use of thermally-controlled fiber Lyot filters [17, 18]. In these studies, the wavelength tuning techniques involve the additional of an intracavity optical filter component in the laser cavity to define the laser operating wavelength. The optical filter component is tuned via strain or temperature to alter the optical filtering characteristics. These strain/thermal tuning methods have limited tuning speed and relies on an intervening optical filter component to define the operating wavelength.

In 2023, a figure-9 mode-locked laser with pump power tuning was reported [19], which does not rely on an intervening intracavity optical filter component to tune the operating wavelength. However, this laser can only achieve a mere 5 nm tuning range and incorporated free space optical components. To date, there are no reports of wide-range L-band tunable MLFLs that combine an all-fiber, all-PM configuration without the need of an intra-cavity optical filter component.

In this work, we report the first pump power-controlled, tunable MLFL in a highly robust all-PM and all-fiber configuration, operating in the soliton regime and generating L-band pulses. Both the mode-locking and the wavelength tuning are facilitated by all-PM nonlinear polarization rotation (NPR) effects. A wide tuning range of 20 nm, from 1568.2 nm to 1588.9 nm, can be achieved by simply adjusting the pump power from 45 mW to 115 mW.

The realization of all-PM NPR lasers involves angle splicing of several PMF segments, exhibiting an enhanced reliability and repeatability over non-PM lasers [20].

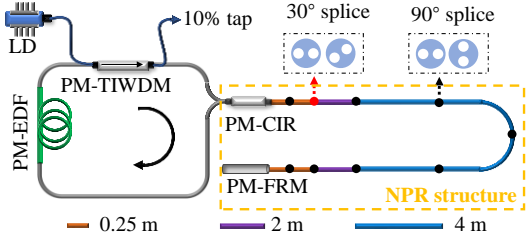

**Fig. 1.** Experimental setup of the wavelength tunable all-PM NPR MLFL. PMF segment length is 0.25 m (orange), 2m (purple), and 4 m (blue).

The experimental setup of our all-PM NPR MLFL is illustrated in Fig. 1, and all the fibers and components are all polarization maintaining. A fast-axis-blocked PM tap isolating wavelength-division multiplexing (PM-TIWDM) acts as a polarizer, a WDM, an isolator, and a 10% output coupler. A 1.4-m-length of PM-EDF (Nufern ESF-7/125) is backward pumped by a 980 nm laser diode (LD) and served as the gain medium. The remaining fibers within the laser cavity are PANDA PM fibers (Fujikura SM15-PS-U25A). A fast-axis-blocked PM circulator (PM-CIR) functions as both a polarizer and an isolator, ensuring unidirectional propagation of the laser in the clockwise direction. The NPR section consists of a PM Faraday rotation mirror (PM-FRM), a PM-CIR, and a 21-m-length of PMF in between. The 21-m-length of PMF is comprising of 10 segments, with the specific length of each segment in meters marked by different colors in Fig.1. The angles at which each adjacent fiber segment is spliced vary as follows: there is one splicing angle of 30°, denoted in red, while the splicing angles for the remaining segments are all 90° (marked in black). The angle splicing is conducted using a splicer (Fujikura FSM-100P), ensuring the loss at each splice point remains below 0.05 dB. The total cavity length is estimated to be 50 m, yielding a net group velocity dispersion of -1.03 ps$^2$ at 1550 nm, indicates operation of the laser is in the soliton regime.

The artificial saturable absorber (SA) mechanism of the NPR structure can be elucidated as follows: From port 1 to port 2 of the PM-CIR, a linear polarized light pulse travel along the slow-axis of the PMF. The light then shifts to the fast-axis of the PMF after the first splicing of 90°. Then the second splicing of 30° leads to the division of the linear light into two orthogonally polarized components along the PMF's principal axes, each carrying different pulse energies. As the two components travels the round-trip over the 21-m-length of PMF, they experience distinct phase shifts due to cross-phase modulation (XPM) and self-phase modulation (SPM). This nonlinear phase accumulation creates different polarization states between the pulse peaks and wings. The FRM rotates the polarization of the two light components by 90°, ensuring they travel equal distance along each axis of the PMF during a complete round trip. This effectively counterbalances the walk-off caused by the birefringence of the PMF [21]. As the reflected light arrives at port 3 of the PM-CIR, it functions as a polarizer, imposing polarization-dependent loss differentially, higher on the pulse wings and lower on the peak. Thus, the NPR structure functions effectively as an artificial SA, facilitating mode-locking.

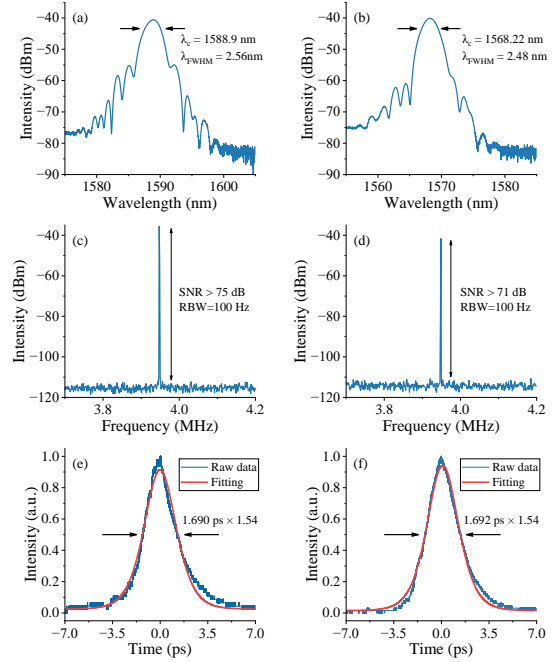

**Fig. 2.** Laser output characteristics: (a), (b) optical spectrum; (c), (d) RF spectrum (RBW,100 Hz); (e), (f) AC trace (after 5 m PMF) at 1588.9 nm and 1568.2 nm, respectively.

The laser self-starts in a multi-pulse mode at a pump power of ~250 mW. Gradually lowering the pump power to ~115 mW induces a transition to a stable single-pulse operation, yielding an average output of ~80 μW at a central wavelength of 1588.9 nm. As the pump power decreases, the central wavelength continuously blue shifts to a shorter wavelength. The laser reaches its shortest wavelength of 1568.22 nm at ~45mW pump. Fig. 2 illustrates the laser's performance at the longest and shortest wavelengths, representing the initial and final points of the entire tuning range. The optical spectra, acquired by an optical spectrum analyzer (OSA, Yokogawa AQ6370D), are depicted in Fig. 2(a) and 2(b). The tuning range initiates at a wavelength of 1588.9 nm, exhibiting a full-width-at-half-maximum (FWHM) spectrum bandwidth of 2.56 nm, and concludes at 1568.2 nm with a FWHM bandwidth of 2.48 nm. To capture the radio frequency (RF) spectrum, a photodetector (New Focus 1611) and an electrical spectrum analyzer (RIGOL RSA3045) are employed. Fig. 2(c) and 2(d) display the RF spectrum with a resolution bandwidth (RBW) of 100 Hz, presenting a signal-to-noise ratio (SNR) of 75 dB at 1588.9 nm and 71 dB at 1568.2 nm, thereby confirming high pulse stability. The 3.9 MHz center frequency aligns with the ~50 m cavity length.

The autocorrelation (AC) trace of the output pulse after a 5 m fiber pigtail, is captured by directly recording the original output with an autocorrelator (Femtochrome FR 103XL) and an oscilloscope (RIGOL MSO8104). The oscilloscope is configured in average mode with 16 times of averaging. Fig. 2(e) and 2(f) depict the FWHM pulse width, estimated under the assumption of a squared hyperbolic secant (sech$^2$) profile, to be 1.69 ps at 1588.9 nm and 1.692 ps at 1568.2 nm, respectively. Given the low pulse peak power, we consider only chromatic dispersion for output pulse propagation in the 5 m fiber pigtail. With a dispersion parameter of 23 ps/(nm·km) near 1.6 μm [22], the FWHM pulse width of the output, after 1 m of PMF propagation, is estimated to broaden by 0.059 ps at 1588.9 nm and 0.057 ps at 1568.2 nm. Therefore, the estimated intracavity FWHM pulse widths are 1.395 ps at 1588.9 nm and 1.407 ps at 1568.2 nm. Corresponding time-bandwidth products are 0.424 and 0.425, indicating the pulses are slightly chirped.

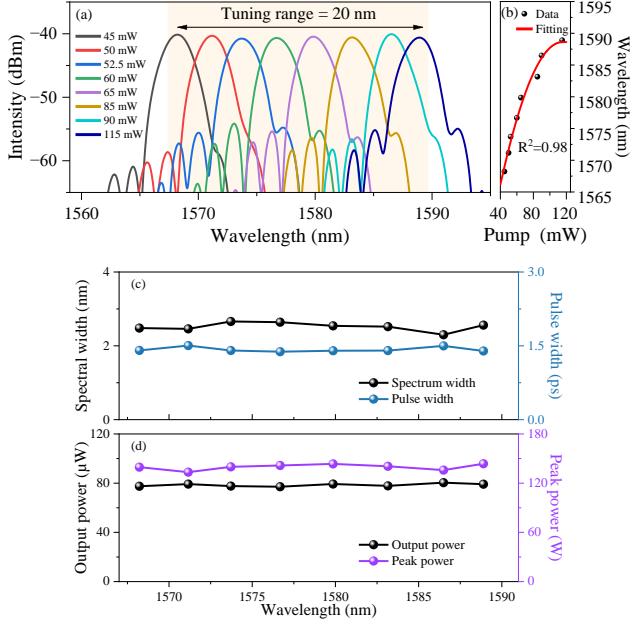

**Fig. 3.** Laser tuning performance. (a) optical spectra at different pump powers; (b) center wavelength versus pump power; (c) FWHM spectral width and corresponding pulse width; (d) output average power and intracavity peak power at different wavelength.

L-band lasing results from a gain profile shift, facilitated by employing a longer standard PM-EDF and lower pump power [20, 23]. The center wavelength-tuning spectra are depicted in Fig. 3(a). The laser maintains mode-locking during this tuning process, and this this 20 nm tuning range can be repeatedly realized by either increasing or reducing the pump power. However, reducing the pump power below 40 mW results in continuous wave lasing. As shown in Fig. 3(b), the quadratic polynomial fitting of the center wavelength against pump power yields an R-square ($R^2$) value of 0.98, indicating a nonlinear correlation between the pump power and the pulse center wavelength. The observed nonlinear correlation likely arises from the uneven gain spectrum and gain profile shifts at different pump powers. The change in pump power nonlinearly affects the pulse peak power, which in turn influences the center wavelength. This is due to the nature of NPR mode-locking, where nonlinear phase accumulation depends on both the pulse peak power and the center wavelength. In Fig. 3(c), the FWHM spectral width and pulse widths at different wavelengths are shown. The spectral width spans from 2.3 nm to 2.66 nm, and pulse widths are estimated to vary from 1.38 ps to 1.5 ps. With the repetition rate variation of about 0.03% during the 20 nm tuning range, we treat the repetition rate as constant when calculating the pulse peak power in the tuning process. Fig. 4(d) displays both the output average power and the estimated intracavity peak power of the pulses, where a standard deviation of 1.4% in average power demonstrates a relative stable pulse output during the tuning process.

The pump power dependent wavelength tuning does not originate from the EDF gain curve shift, while the gain peak typically red shift to longer wavelengths with decreasing pump power [24, 25]. However, in our laser, the center wavelength experiences a blueshift to shorter values as the pump power reduces. The wavelength tuning is likely due to the mode-locking mechanism, which is the all-PM NPR in our laser. Wavelength tuning is characterized by a shift from mode-locking at one wavelength to another. This implies that both the original and new wavelengths meet the criteria for effective mode-locking by the all-PM NPR, which is based on the nonlinear polarization rotation. Given that the rotation depends on the overall phase difference between the two orthogonal light components, suggesting a consistent or similar overall phase difference is maintained across both the original and new wavelengths. For a simplified analysis, we assume the overall phase difference between the two light components is preserved throughout the tuning process.

The nonlinear phase shifts of the two polarized components, accumulated via XPM and SPM, are denoted as follows [26]

$$\varphi_s = \gamma L(|E_s|^2 + \frac{2}{3}|E_f|^2)$$
$$\varphi_f = \gamma L(|E_f|^2 + \frac{2}{3}|E_s|^2) \quad (1)$$

where, nonlinear parameter $\gamma = 2\pi n_2/(\lambda A_{eff})$, $n_2$ is nonlinear refractive index, $A_{eff}$ is effective mode area. $L$ represents the fiber length. $E_s$ and $E_f$ denote the electrical amplitudes of the light components along the slow and fast axes, respectively. The ratio of $E_s$ to $E_f$ is determined by the 30° angle splicing [27]. In our reflected all-PM NPR setup, both the two light components travel equal distances along the PMF axes, resulting in the same value of $L$ and identical linear phase shifts for each component. Therefore, any difference in their overall phase is solely due to variations in the nonlinear phase shift. The overall phase difference can be expressed as:

$$\Delta\varphi = \varphi_s - \varphi_f = \frac{2\pi L n_2}{3}\frac{(|E_s|^2 - |E_f|^2)}{\lambda A_{eff}} \quad (2)$$

Given the 20 nm spectrum tuning range, the $n_2$ varies minimally, allowing us to treat the factor ($2\pi L n_2/3$) on the right-hand side as constant during the tuning process. Under the assumption that the overall phase difference ($\Delta\varphi$) remains constant throughout the wavelength tuning process, the factor (($|E_s|^2-|E_f|^2)/\lambda A_{eff}$) should also remain constant. Given that $A_{eff}$ is proportion to $\lambda^2$ [28], in our analysis, we consider $\lambda A_{eff}$ to be proportional to $\lambda^3$. Therefore, a reduction in pump power leads to a decrease in pulse peak power, subsequently causing a reduction in ($|E_s|^2-|E_f|^2$). This in return, results in a corresponding decrease in $\lambda$, ensuring the factor $\Delta\varphi$ remains consistent throughout the tuning process. This mechanism likely accounts for the phenomenon observed in the experiment, where reduced pump power correlates with a blue shift of the center wavelengths.

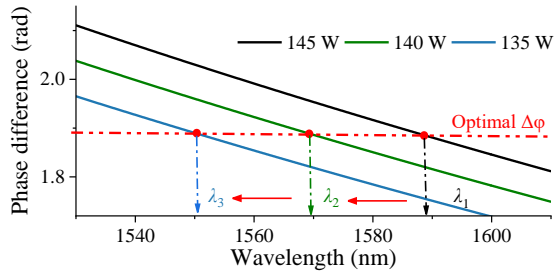

**Fig. 4** Simulated overall phase difference at different pulse peak power; the red dashed line marks the assumed optimal phase difference.

Fig. 4 shows the simulated overall phase difference at various pulse peak powers. The key parameters include a splicing angle of 30°, and a fiber length of 42m. The nonlinear parameter $\gamma$ is treated as inversely proportional to $\lambda^3$, set at 2 $W^{-1}km^{-1}$ at 1.55 μm [22]. To reflect experimental conditions, the simulation's pulse peak powers are set near the experimental values at the initial and end wavelengths in the tuning process, being 145 W and 140 W, respectively, and includes an additional setting of 135 W for demonstration. The $\lambda_1$ corresponds to the initial experimental wavelength at 1588.9 nm, and the optimal phase difference (red dashed line) in simulation is determined at this wavelength when peak power is 145 W. Under the assumption of a consistent phase difference ($\Delta\varphi$) is kept between the original and new wavelengths during tuning, the simulation yields a 19.5 nm tuning range. This is calculated by $\lambda_1$ minus $\lambda_2$ as the pulse peak power decreases from 145 W to 140 W, closely approximating the 20.7 nm tuning range observed in the experiment. The center wavelength shifts from $\lambda_1$ to $\lambda_3$ as pulse peak power reduces from 145 W to 135 W, clearly indicating a blue shift trend in the center wavelength as the pulse peak power decreases.

In summary, we present the first all-PM, all-fiber, pump power-controlled, wavelength tunable L-band MLFL. Utilizing all-PM NPR for both mode-locking and wavelength tuning, a 20 nm tuning range (1568.2 nm tp 1588.9 nm) is achieved by adjusting the pump power from 45 mW to 115 mW. Contrasting with conventional wavelength tuning methods that rely on optical filters, loss control, or birefringence adjustments, our laser employs a pump power tuning method, realizing a simple and cost-effective design with enhanced robustness, reliability, repeatability and speed.

**Funding.** Japan Society for the Promotion of Science (18H05238, 22H00209, 23H00174); Core Research for Evolutional Science and Technology (JPMJCR1872).

**Disclosures.** A US provisional patent application related to the work presented in this paper has been filed.

**Data Availability.** Data underlying the results presented in this paper are not publicly available at this time but may be obtained from the authors upon reasonable request.

**Acknowledgements**. Sze Yun Set thanks Mr. Hideru Sato for personal donation partially supporting this research work.